\documentstyle[preprint,prc,aps,psfig]{revtex}

\tightenlines

\begin{document}

\title{Fusion cross sections at deep subbarrier energies} 
\author{K. Hagino$^{1,2}$, N. Rowley$^3$, and 
M. Dasgupta$^4$}
\address{$^1$Yukawa Institute for Theoretical Physics, Kyoto
University, Kyoto 606-8502, Japan }
\address{$^2$Institut de Physique Nucl\'eaire, IN2P3-CNRS, \\ 
Universit\'e Paris-Sud, F-91406 Orsay Cedex, France}
\address{$^3$Institut de Recherches Subatomiques, 
UMR7500, IN2P3-CNRS/Universtit\'e Louis Pasteur, BP28, F-67037 
Strasbourg Cedex 2, France}
\address{$^4$
Department of Nuclear Physics, Research School of Physical 
Sciences and Engineering, Australian National University, 
Canberra ACT0200, Australia}

\maketitle

\begin{abstract}

A recent publication reports that  
heavy-ion fusion cross sections at extreme subbarrier energies 
show a continuous change of their logarithmic slope with decreasing 
energy, resulting in a much steeper excitation function compared 
with theoretical predictions. 
We show that the energy dependence of this slope is 
partly due to the asymmetric shape of the Coulomb barrier, that is 
its deviation from a harmonic shape. We also point out that the 
large low-energy slope is consistent 
with the surprisingly large surface diffusenesses 
required to fit recent high-precision fusion data. 

\end{abstract}

\pacs{PACS numbers: 25.70.Jj,24.10.Eq}

The primary ingredient in any nuclear reaction calculation 
is the nucleus-nucleus potential, consisting of the 
repulsive Coulomb interaction and an attractive nuclear part. 
Although the Coulomb term $V_C(r)$ is well-known, 
there are large ambiguities 
in the nucleus-nucleus potential $V_n(r)$, and many attempts 
have been made to extract 
information on this quantity from experimental data for 
heavy-ion reactions. While elastic and inelastic scattering 
are sensitive mainly to the surface region of the nuclear potential, the 
fusion reaction is also relatively sensitive to the inner part. They 
thus provide complementary sources of information. 

In heavy-ion reactions, strong channel coupling effects 
(due to collective inelastic excitations of the colliding 
nuclei and/or transfer processes) significantly modify 
the landscape of potential energy surface, replacing the uncoupled 
single barrier with a distribution of barriers 
\cite{RSS91,DHRS98,BT98,L95}. 
In order to extract the nucleus-nucleus potential 
from heavy-ion fusion reactions, 
it is therefore advisable to use either high-energy fusion data where the 
barrier penetrability is essentially unity for all the distributed barriers, 
or very low-energy data where only the lowest barrier contributes 
to the cross section. 
Of these, the low-energy data probably provide 
cleaner information since the high-energy data may be 
complicated by competing reaction processes such as deep-inelastic 
scattering. 

A recent paper~\cite{JER02} has reported an attempt to measure 
the fusion cross section $\sigma$ for the $^{60}$Ni + $^{89}$Y system 
at deep subbarrier energies, down to the 10$^{-4}$ mb level. 
The authors of Ref.~\cite{JER02} used the Wong fusion
formula~\cite{W73} to analyse their data and 
showed that the experimental cross section exhibited 
an abrupt decrease at extreme subbarrier energies. They also analysed 
the data in terms of the logarithmic slope, defined by 
$L(E)=d(\ln(\sigma E))/dE$, and showed that this quantity 
exhibited a continuous increase with decreasing energy, 
in contrast to the theoretical slope which approached a constant value. 
They also found similar behavior in a few other systems found in the 
literature, including the $^{58}$Ni + $^{58}$Ni and 
$^{90}$Zr + $^{92}$Zr reactions. 

The main part of the analysis in Ref.~\cite{JER02} 
relied on the Wong formula as a reference. 
A natural question is whether this 
formula, based on a parabolic approximation to the Coulomb 
barrier, is adequate at deep subbarrier energies~\cite{RM91}. 
It was claimed in Ref.~\cite{JER02} that the Wong formula leads to 
fusion cross sections similar to those obtained with 
the coupled-channels approach 
for the $^{58}$Ni + $^{58}$Ni system. However, the former was simply a
fit to the latter with parameters which had no 
physical connection to the
potential used in the coupled-channels calculations. 

The aim of this paper is to reanalyse critically the $^{58}$Ni 
+ $^{58}$Ni reaction with an exact one-dimensional-potential
calculation 
as well as with coupled-channels calculations~\cite{HRK99} 
and show that the 
Wong formula is indeed unreliable at 
very low energies. This is particularly so for a quantity 
such as the logarithmic slope which accentuates the 
energy dependence of the cross section. 
We also discuss the findings of Ref.~\cite{JER02} in connection 
with the problem of the large surface diffusenesses of the nuclear 
potential for 
subbarrier fusion, discussed for some time in the 
literature~\cite{L95,HDG01,NMD01}. 

Let us first discuss the validity of the parabolic approximation to the 
potential. Fig.~1 shows the nucleus-nucleus potential for the $^{58}$Ni 
+ $^{58}$Ni system (solid line), along with its parabolic approximation
(dashed line) 
\begin{equation}
V(r)\sim B - \frac{1}{2}\mu \omega^2 (r-R)^2,
\label{parabolic}
\end{equation}
where $B$ and $R$ are the barrier height and position, 
respectively. Here $\mu$ is the reduced mass of the system and $\omega$ is 
the barrier ``curvature'' given by $\omega^2=-V''(R)/\mu$. We use a 
Woods-Saxon nuclear potential with $V_0=$ 160~MeV, $r_0$ = 1.1~fm and 
$a$= 0.65~fm. On the inside, the nuclear 
potential varies relatively rapidly, while on the outside the Coulomb 
potential varies slowly, resulting in an asymmetric barrier shape.
The deviation from the parabolic 
approximation~(\ref{parabolic}) becomes larger as 
the energy goes down and one 
expects this approximation to break down at energies 
well below the barrier. 
It was shown in Ref.~\cite{RM91} that the parabolic 
approximation is adequate only for $|r-R| \le a$, 
that is for  
incident energies within $\mu\omega^2a^2/2$ of the barrier height. In the 
present example $\mu\omega^2a^2/2$ = 2.62~MeV, and it is evident that the 
parabolic approximation is valid only in a relatively small range of 
energies near the barrier top. 

An analytic formula for the fusion cross section for the parabolic 
barrier (\ref{parabolic}) was derived some time ago by Wong~\cite{W73}: 
\begin{equation}
\sigma(E)=\frac{\hbar\omega}{2E}
R^2\ln\left[1+e^{2\pi(E-B)/\hbar\omega}\right].
\label{Wong}
\end{equation}
The upper panel of Fig.~2 compares this formula with 
the fusion cross section obtained by numerically solving the 
Schr\"odinger equation with the true potential. 
No coupling is included in these calculations. 
As we saw in Fig.~1, the parabolic approximation underestimates the 
barrier thickness in the tunnelling region, and thus overestimates the 
penetrability at low energies. The bottom panel of 
Fig.~2 shows the logarithmic slopes $L(E)$.
Eq.~(\ref{Wong}) yields a slope that is
constant at low energies and is given by 
\begin{equation}
L(E)\sim 
\frac{d}{dE}\ln\left[\frac{\hbar\omega}{2}R^2e^{2\pi(E-B)/\hbar\omega}
\right]
= \frac{2\pi}{\hbar\omega}.
\label{slope}
\end{equation}
On the other hand, the slope computed from the exact 
results shows a continuous increase with decreasing incident energy 
(solid line). This is reminiscent of the experimental findings of 
Ref.~\cite{JER02}. 

At low energies, the logarithmic slope is related 
to the s-wave barrier penetrability $P_0$ 
by $L(E)=d\ln(P_0(E))/dE$. In the WKB approximation, the penetrability is 
given by 
\begin{equation}
P_0(E)= e^{-2S(E)/\hbar} = 
\exp\left[-2\int^{r_2}_{r_1}dr\,\sqrt{2\mu(V(r)-E)/\hbar^2}\right]
\end{equation}
at energies well below the barrier. Here, $r_1$ and $r_2$ are the inner 
and the outer turning points, respectively. Defining $\Delta(E)$ 
as the difference between the true action integral $S(E)$ 
and its value in the quadratic approximation 
we have (ingoring an unimportant constant factor):
\begin{equation}
S(E)=\int^{r_2}_{r_1}dr\,\sqrt{2\mu(V(r)-E)}
=\frac{\pi}{\omega}(B-E) + \frac{\hbar}{2}\Delta(E). 
\end{equation}
Since the Coulomb barrier $V(r)$ has a non-symmetric shape, 
$\Delta(E)$ increases as the energy decreases and the 
logarithmic slope $L(E)=2\pi/\hbar\omega - d \Delta(E)/dE$ is 
always larger than $2\pi/\hbar\omega$. Furthermore, one 
can show that the second derivative of this action integral 
is a positive quantity and thus 
$L(E)$ is a decreasing function of $E$. 
For example, this is the case for the sharp-cut 
potential,  $V_n(r)=\left[-V_C(r)-V_0\right]\,\theta(R_0-r)$, 
for which the action integral can be evaluated 
analytically~\cite{Gamow}. 
These facts are consistent with the numerical result shown in the 
lower panel of Fig.~2 as well as with the experimental findings discussed in 
Ref.~\cite{JER02}. We thus conclude that the continuous increase of 
the logarithmic slope with decreasing energy is not in itself
evidence of anomalous behavior of the fusion cross section at 
very low energies, as claimed in Ref.\cite{JER02}. 

We now discuss the relation between the logarithmic slope $L(E)$ 
and the surface property of the nuclear 
potential. For scattering processes, it seems well accepted that
the surface diffuseness parameter $a$ should be around 0.63~fm 
if $V_n$ is parametrised by a Woods-Saxon form 
\cite{CW76,CPR96,SAC01}. In marked contrast, recent high-precision 
fusion data suggest that a much larger 
diffuseness, between 0.8 and 1.4~fm, is needed 
to fit the data~\cite{HDG01}. This is not just for particular 
systems but seems to be a rather general 
result\cite{L95,NMD01,M99,M01,SACN95,SC01}. 
Note that fusion depends strongly on the potential on both sides 
of the barrier, in contrast to the elastic scattering which 
depends mainly on the potential on the outside. 
At high energies, the fusion cross section changes with the diffuseness 
due to the way the position and height of the $l$-dependent barrier 
change with increasing $l$. At lower energies, 
the main effect comes from the overall width of the barrier. 
A large diffuseness seems to be derirable in both these 
respects~\cite{HDG01}. 

For a fixed value of the 
barrier height $B$, the barrier curvature $\hbar\omega$ is 
approximately proportional to $a^{-1/2}$~\cite{RM91}. 
Eq.~(\ref{slope}) then indicates that 
the logarithmic slope $L(E)$ is roughly proportional to $a^{1/2}$. 
The large experimental slope found in Ref. 
\cite{JER02} may therefore be another indication of the large surface 
diffusenesses already noted in heavy-ion fusion. 
In order to assess this, we perform the exact coupled-channels 
calculations for the $^{58}$Ni + $^{58}$Ni 
reaction using the computer code {\tt CCFULL}~\cite{HRK99} 
with different values of the surface diffuseness. 
This code uses the isocentrifugal approximation to reduce 
the dimensionality of the coupled-channels equations 
(see Ref.~\cite{HRK99} for 
details) but we have checked that this is still valid 
at energies well below the Coulomb barrier. 
In the 
calculations, we include the double quadrupole-phonon 
excitations in both the projectile and target nuclei. A similar 
coupling scheme successfully explained the experimental fusion 
cross section and barrier distribution for the very similar
$^{58}$Ni + $^{60}$Ni system~\cite{SACN95}. The dynamical quadrupole 
deformation parameter $\beta_2$ for the Coulomb coupling 
is estimated to be 0.177 from the 
experimental $B(E2)$~\cite{ND80} with the radius parameter 
$r_{\rm coup}=1.2$~fm. We require a somewhat larger value of
$\beta_2$ = 0.261 (with $r_{\rm coup}=1.06$~fm) for the nuclear 
coupling in order to fit the data. 
The fusion reaction often requires a radius parameter of
around 1.06~fm, smaller than the usual value of around 1.2~fm,
used to extract a 
deformation parameter from the electromagnetic transition probability. 
This results in a larger deformation parameter as well as in
a larger deformation length $\beta \cdot r_{\rm coup}$. Although the 
Coulomb-coupling hamiltonian is independent of the value of the radius 
parameter to be used, the nuclear coupling term depends on it through 
the combination $\beta \cdot r_{\rm coup}$. 
Therefore, this problem may also 
be related to the parametrization of the nucleus-nucleus 
potential and thus to the large surface diffuseness problem, 
though the value of $r_{\rm coup}$ = 1.06~fm should be 
reasonable for finite nuclei with a diffuse surface~\cite{BR74}. 

In Fig.~3, we show the dependence of the fusion cross section 
(upper panel) and of the 
logarithmic slope (lower panel) on the surface 
diffuseness parameter $a$ for the $^{58}$Ni + $^{58}$Ni reaction. 
The figure also includes the experimental data~\cite{BBE81} for 
comparison. 
The experimental slope was computed using point-difference formulae 
with both two and three successive data points. 
The dotted line is the result 
with the nuclear potential shown in Fig.~1, that is 
with $a$=0.65~fm, while the dashed line is obtained with the 
potential parameters $V_0=$ 195~MeV, $r_0$ = 0.94~fm and $a$= 1.0~fm. 
The former leads to a cross section whose slope is not steep 
enough to account for the experimental data at energies below the 
barrier. As a consequence, the logarithmic slope $L(E)$ is underestimated 
at these energies, as in Ref.~\cite{JER02}. On the other hand, 
the potential with $a$=1.0~fm improves the agreement considerably 
both for the cross section and the logarithmic slope. 
We also include in the figure a calculation with $a=1.3$~fm (solid line). 
This further improves the 
fit to the logarithmic slope, although it somewhat worsens the fit 
to the cross section itself at incident energies around 97~MeV. 
(We have confirmed that none of these results 
depends on the value of $r_0$ 
as long as $V_0$ is adujusted so that the barrier 
height remains unchanged.) 
Clearly, 
the experimental data favor a large value of the surface diffuseness, 
as in many other systems in the literature. 

In summary, the ``unexpected'' behavior of heavy-ion fusion cross sections 
at extreme subbarrier energies claimed in Ref.~\cite{JER02} has two causes. 
One is the use of the Wong formula which is inadequate at energies 
far below the barrier. The exact numerical calculation is vital in 
discussing the fusion cross section and especially 
the logarithmic slope $L(E)$ at low energies. We pointed out that the 
exact calculation shows a similar energy dependence of the logarithmic 
slope as in the experimental data even without coupling. 
The other reason for this apparent anomaly is the use of 
a diffuseness parameter which is widely used in calculations 
for scattering processes, that is $a\approx$ 0.63~fm. 
This potential leads to fusion cross sections whose logarithmic 
slope is much smaller than for the experimental data at deep subbarrier 
energies. 
If such a calculation is 
used as a reference, the experimental data may appear to fall much 
more steeply than expected~\cite{JER02}. 
However, if one
uses a larger value of the diffuseness parameter in the phenomenological
potential, 
the data can be reproduced within the present 
coupled-channels framework. The need for a large diffuseness 
to describe the fusion process 
has also been found consistently in other systems. 
However, the reason for the large differences in diffuseness parameters 
extracted from scattering and from fusion analyses remains 
an open problem. 
In particular, it is still not clear whether 
a large surface diffuseness reflects the true nature of the potential 
or simply mocks up other effects which cause a rapid decrease 
of fusion at low energies. 
In this context, we mention 
that neither the double-folding potential~\cite{HDG01} (which is 
usually much deeper and 
narrower than the Woods-Saxon) 
nor the geometrical corrections to the coupling potential~\cite{GDH02} 
seem to resolve this problem. 

More experimental 
and theoretical studies of fusion at deep 
subbarrier energies are needed to improve our understanding 
of this process which may be especially important in astrophysical 
fusion reactions. Isotopic dependences may also be of interest,
particularly for exotic nuclei whose surface properties may be
modified by the presence of weakly bound nucleons.

\bigskip

K.H. thanks the Nuclear Theory Group at the IPN Orsay for its hospitality,  
and the Kyoto University Foundation for financial support. 
M.D. acknowledges financial support from the Australian Research Council.

\begin{figure}
  \begin{center}
    \leavevmode
    \parbox{0.9\textwidth}
           {\psfig{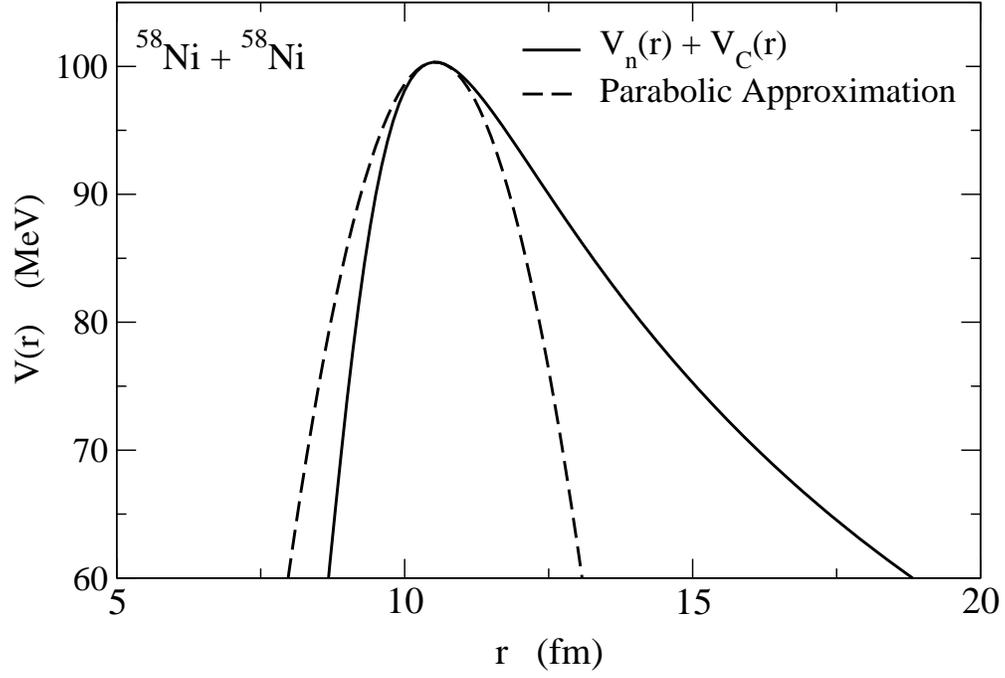}}
  \end{center}
\protect\caption{
The nucleus-nucleus potential for the $^{58}$Ni +  $^{58}$Ni  
reaction. The solid line is obtained with a Woods-Saxon 
nuclear potential with parameters $V_0$=160~MeV, $r_0$=1.1~fm, 
and $a=$ 0.65~fm. The dashed line shows the quadratic expansion 
of the potential around the barrier position. 
}
\end{figure}

\begin{figure}
  \begin{center}
    \leavevmode
    \parbox{0.9\textwidth}
           {\psfig{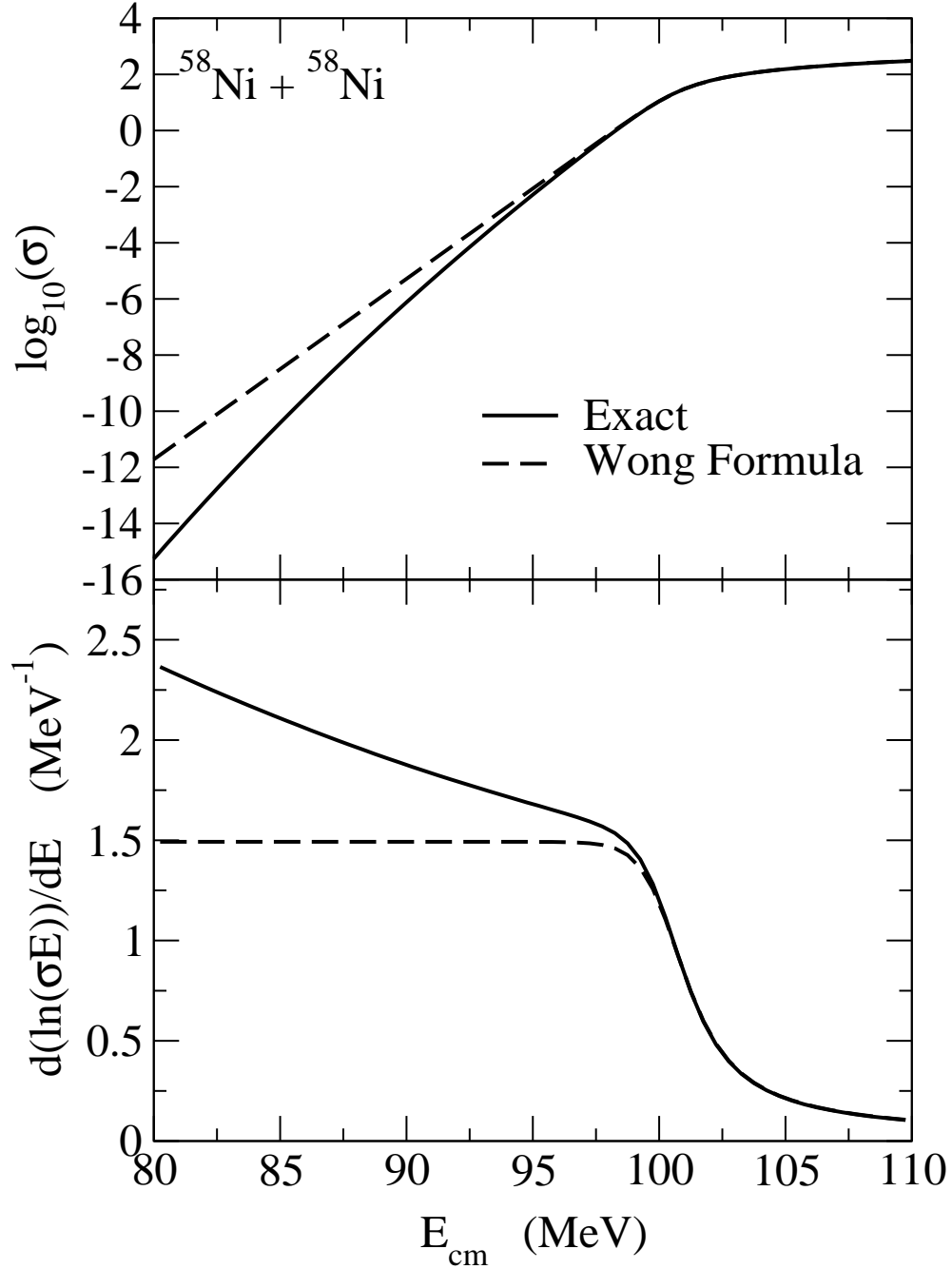}}
  \end{center}
\protect\caption{
The validity of the Wong formula (2) for the fusion cross section 
for the $^{58}$Ni +  $^{58}$Ni system. The upper panel shows 
the fusion cross section $\sigma$ (in mb) on a logarithmic scale, 
while the lower panel shows the logarithmic slope 
$L(E)=d(\ln(\sigma E))/dE$. The solid and dashed lines 
denote the exact numerical results and the Wong cross section, 
respectively.
}
\end{figure}

\begin{figure}
  \begin{center}
    \leavevmode
    \parbox{0.9\textwidth}
           {\psfig{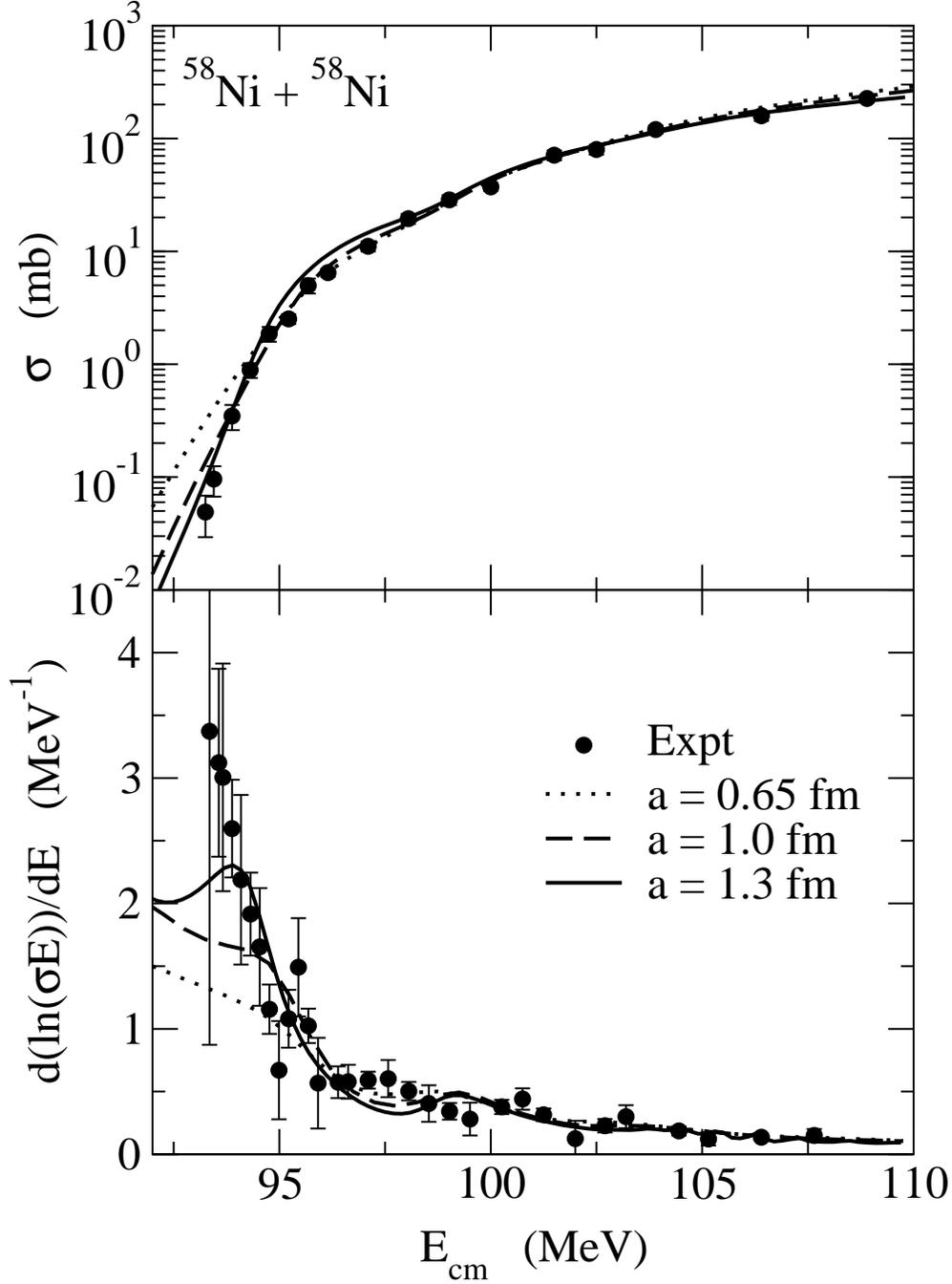}}
  \end{center}
\protect\caption{
Dependence of the fusion cross section (upper panel) and the 
logarithmic slope (lower panel) on the surface diffuseness 
parameter $a$ 
for the $^{58}$Ni +  $^{58}$Ni reaction. The dotted, dashed and 
solid lines are coupled-channels results using 
diffuseness parameters of 0.65~fm, 1.0~fm and 1.3~fm, respectively. 
The double quadrupole-phonon exitations in both the projectile 
and target are taken into account. Experimental data 
are from Ref. {\protect\cite{BBE81}}. 
}
\end{figure}


\begin{references}
\bibitem{Gamow}In the limit of $R_0\to 0$, the action integral for this 
model is given by $\pi\eta(E)$, where $\eta(E)=Z_PZ_Te^2/\hbar v$ is the 
Sommerfeld parameter. For this, $S'(E)$ and $S''(E)$ are proportional to 
$-E^{-3/2}/2$ and $+3E^{-5/2}/4$, respectively.
\end{references}
\end{document}